# 9. More than programming? The impact of AI on work and skills

Toby Walsh
https://orcid.org/0000-0003-2998-8668

**Abstract**
This chapter explores the ways in which organisational readiness and scientific advances in Artificial Intelligence have been affecting the demand for skills and their training in Australia and other nations leading in the promotion, use or development of AI. The consensus appears that having adequate numbers of qualified data scientists and machine learning experts is critical for meeting the challenges ahead. The chapter asks what this may mean for Australia's education and training system, what needs to be taught and learned, and whether technical skills are all that matter.

> "In my opinion, ignoring AI is like ignoring blogging in the lates 1990s, or social media circa 2004, or mobile in 2007. Very quickly, some degree of facility with these tools will be increasingly essential for all professionals, a primary driver for new opportunities and new jobs. Developing skills and competencies in it now will yield benefits for years to come.
> It's also true that the changes AI will bring will have negative impacts as well as positive ones. Previous technology revolutions disrupted specific subgroups, like craftsmen whose production was replace by factories – or, more recently, factory workers who lost their jobs to increased automation.
> Now, knowledge workers are also facing these challenges. While I strongly believe that these new AI tools will create new jobs and new industries, along with great economic benefits and other quality-of-life gains, they
> will also eliminate some jobs, both blue- and white-collar."
> (Hoffman & GPT4, 2023, p.110-111)

## 9.1 Introduction

It seems certain that artificial intelligence (AI) will have a large impact on economy and society. For instance, one study estimates that it will grow the world's economy by around 15 per cent or so, contributing over $15 trillion annually in inflation adjusted terms (Rao & Verweij, 2017). Such disruption will not be without pain. One of the greatest fears about AI is the impact it will have on work. Will it eliminate many jobs? Will workers who embrace AI be more productive and replace those workers who do not? For jobs that are not replaced by AI, how will the skills that workers need evolve? For the new jobs that AI creates, what skills will be needed? And will AI be a net positive, creating more jobs than it destroys? Or will it be a net negative?

It is clear already that the impacts will not be even. Some countries will be impacted greater than others. Even within a country, impacts will not be even. Some sectors of the economy will be more severely disrupted than others. Predicting where those impacts will fall and what they will be is not easy. It isn't as simple as the blue collar workers doing manual work who will be replaced by robots and white collar workers doing cognitive work who will be saved.

There are, unfortunately, blue collar workers doing jobs too poorly paid for it to be economically viable to replace. There are also blue collar workers doing jobs that robots cannot do. Plumbers and electricians are, for example, likely safe from automation for a long time. And some blue collar jobs may be invulnerable to technological disruption even when AI could, in theory, do them. We will likely always value the artisan over the mass produced so, in opposition to what Reid Hoffman argued in the opening quotation, jobs such as cabinet maker may remain even when machines could in theory do the work. On the other hand, there are white collar workers who are perhaps less confident today that their jobs are safe from automation than they were a decade ago. For instance, graphic designers were perhaps not too concerned about their jobs until image to text tools like Stable Diffusion and DALL-e arrived hinting at a future where a lot of graphic design might be automated.

Computer programmers, especially those writing (AI) software might be more confident than graphic designers that that their skills will remain in high demand. But even they cannot be certain. The irony is that machine learning is getting computers to program themselves. The largest bits of software constructed today are large neural networks, with billions of parameters that are set by gradient descent and not by any human. In addition, even when limiting our attention to more conventional software, AI is changing the nature of computer programming. Large language models can easily be trained not to produce natural language but computer code. This should not be surprising as computer languages are more regular than natural language. Large language models trained on computer code can greatly improve the productivity of a competent programmer. One able programmer with such a tool might therefore be able to do the work of several human programmers. Will this mean we need fewer computer programmers in the future?

## 9.2 Technological unemployment

The fear that technology would disrupt employment is an old one. In 1930, John Maynard Keynes warned:

> "a new disease of which some readers may not have heard the name, but of which they will hear a great deal in the years to come: namely, technological unemployment. This means unemployment due to our discovery of means of economising the use of labour outrunning the pace at which we can find new uses for labour."
> (Keynes, 1930, p. 325)

Economists can often be wrong so perhaps it should not be surprising that this prediction has not, at least yet, come true. Unemployment did shortly become a major problem. but the reason was not technological but financial. The Great Depression led to a rapid rise in unemployment that lasted a decade. Today, however, unemployment has fallen and is at historic low levels in most countries. This is despite the world's population being at historic high levels. Work hasn't ended, though many of us are working fewer hours.

Nevertheless, fears about technological unemployment have continued to grow since then. In 1949, Alan Turing put it in very plain terms[1] in an interview with the Times newspaper:

> "This is only a foretaste of what is to come, and only the shadow of what is going to be. We have to have some experience with the machine [the Manchester Mark 1 computer] before we really know its capabilities. It may take years before we settle down to the new possibilities, but I do not see why it should not enter any one of the fields normally covered by the human intellect, and eventually compete on equal terms" (Turing, 1947).

If machines can compete with humans, then what chance is there for humans? It is hard to understand how Turing can suggest machines would compete on equal terms with humans. Machines don't need to rest, or be paid. These are surely unfair advantages and unequal terms?

Three years after Turing's comment, the famous economist Wassily Leontief wrote, somewhat optimistically, about how the working week and income distribution would permit the economy to adapt to technological advances (Leontief, 1952). He used horse labour as an example of the threat posed to human labour by technological change. Following the invention of the railroads and the telegraph, the role of horse labour in the US economy actually increased. The equine population grew sixfold between 1840 and 1900 to more than 21 million horses and mules as the US grew and prospered. In 1900, horses might therefore have felt safe from technological change. While their job transporting people and messages between towns and cities had started to disappear, other jobs had arrived to replace this.

Horse weren't to know that this good fortune was to be short lived. The invention of the internal combustion engine rapidly reversed this trend. The US population grew larger, and the nation became richer. But horses began to disappear from the labour market. By 1960, there were just three million horses in the US, a decline of nearly 90 percent. Economists debating the future role of horse labour in the economy in the early 1900s might have predicted that, just as in the past, new jobs for horses would emerge in areas enabled by the new technologies. They would have been very wrong.

These early worries about technology unemployment came to a head in March 1964. President Lyndon Johnson received a short but alarming memorandum from the Ad Hoc Committee on the Triple Revolution (Agger et al., 1964). The memo was signed by luminaries including Nobel Prize winning chemist Linus Pauling, Scientific American publisher Gerard Piel, and Gunnar Myrdal who was to go on to win the Nobel Prize in Economics. The memo warned that technology would soon create mass unemployment. It predicted that automation and computers were set to change the economy in as fundamental a way as the industrial revolution changed the agricultural era before it, and that this revolution would occur at a speed never witnessed before.

---

[1] Britain's 50 Pound bank note includes Alan Turing's portrait along with the first part of this quotation: "This is only a foretaste of what is to come, and only the shadow of what is going to be."

In absolute terms, the memorandum (like Keynes before it) was wrong. There has not been mass unemployment. In fact, since 1964, the US economy has added over 70 million new jobs. But computers and automation have radically changed the jobs that are available, the skills those jobs require, and the wages paid for those jobs. And it is very unlikely that we have got to the end point yet. Like the example of horses at the start of the industrial revolution, we should be cautious about extrapolating forwards from today. There are early warnings of more troubling times ahead.

In 2015, for example, 22 per cent of men in the US without a college degree aged between 21 to 30 had not worked at all during the prior twelve months. Twenty-something male high-school graduates used to be the most reliable cohort of workers in America. They would leave school, get a blue-collar job and work at it till their retirement some forty or more years down the road. Today over one in five are out of work. The employment rate of this group has fallen 10 percentage points. And this appears to have triggered cultural, economic, and social decline. Without jobs, this group is less likely to marry, to leave home, or to engage politically (Wilson, 1987). The future for them looks rather bleak. If they cannot get on the employment ladder, are they going to be forever without a decent job?

**9.3 Machine learning predictions**

A number of studies have tried to quantify the impact more precisely. One of the most widely reported was a study out of the University of Oxford (Frey & Osborne, 2013). This report famously predicts that 47 per cent of jobs in the US are under threat of automation in the next two decades or so. Studies for other countries like Australia have reached broadly similar conclusions. Most recently, similar studies are starting to appear about the impact of generative AI on jobs despite concerns about the methodology of the original Oxford study (e.g. Hatzius et al., 2023).

Ironically, the writing of the 2013 Oxford report was itself partially automated. The authors used machine learning to predict precisely which of 702 different job types could be automated. They used machine learning to train a classifier, a program to predict which jobs would be automated. They first fed the program with a training set, 70 jobs that they had labelled by hand as automatable or not. The program then predicted whether the remaining 632 jobs could be automated. Even the job of predicting which jobs will be automated in the future has already been partially automated.

As with any machine learning problem, the predictions of the classifier depend critically on the training data. The training set of 70 out of the 702 different jobs was classified by hand. The classification was binary: at risk of automation, not at risk of automation. Some of the jobs classified likely fell in-between. For instance, one job which they classified as at risk of automation was accountant and auditor. They are certainly parts of being an accountant and auditor that will be automated in the next few decades. But it is very doubtful that all parts of the job of being an accountant or auditor will disappear.

In total, the hand classified training set had 37 of the 70 jobs at risk of automation. That is, over half of their training data – provided as input to the classifier – were jobs said to be risk of automation. Not surprisingly then, the output of the classifier was a prediction that around half the full set of 702 jobs were at risk of automation. One might expect if their training set had been more cautious, say labelling only one in four jobs in the training set at risk of automation, then their overall prediction on the full set of jobs would have been equally cautious.

To explore this, I ran a survey of my own (Walsh, 2018). I asked 300 experts in AI and Robotics to classify which of the jobs in the training set were at risk of automation in the next two decades. I also asked the same questions of nearly 500 non-experts, members of the public who read an article I wrote about advances in poker bots. The non-experts agreed almost exactly with the classifications in the training set. But the experts in AI and Robotics were significantly more cautious. They predicted around 20 per cent fewer jobs were at risk of automation. This would translate into a significant reduction in the predicted number of jobs at risk of automation.

Even if you agree with all the assumptions and predictions of the Frey and Osborne report, you cannot conclude that half of us will be unemployed in a couple of decades. The report merely estimates the number of jobs that are potentially automatable over the next few decades. There are many reasons why this will not translate into 47 per cent unemployment.

First, the report merely estimated the number of jobs that are susceptible to automation. Some of these jobs won't be automated in practice for economical, societal, technical and other reasons. For example, we can pretty much automate the job of an airline pilot today. Indeed, most of the time, a computer is flying your plane. But society is likely to continue to demand the reassurance of having a pilot on board for some time to come even if they are just reading their iPad most of the time.

Second, we also need to consider all the new jobs that technology will create. For example, we don't employ many people setting type anymore. But we do employ many more people in the digital equivalent, making web pages. Of course, if you are a printer and your job is destroyed, it helps if you're suitably educated so you can reposition yourself in one of these new industries. There is sadly no fundamental law of economics that requires as many new jobs to be created by new technologies as destroyed. It happens to have been the case in the past. But as horses have discovered over the last century, it is not necessarily the case with all new technologies.

Third, some of these jobs will only be partially automated, and automation may in fact enhance our ability to do the job. For example, there are many new tools to automate scientific experiments: gene sequencers that can automatically read our genes, mass spectrometers that can automatically infer chemical structure, and telescopes that can automatically scan the skies. But this hasn't put scientists out of a job. In fact, more scientists are alive and doing science today than have ever lived in the history of civilisation. Automation has lifted their productivity. Scientific knowledge is simply discovered faster.

Fourth, we also need to consider how the working week will change over the next few decades. Most countries in the developed world have seen the number of hours worked per week decrease significantly since the start of the industrial revolution. In the U.S, the average working week has declined from around 60 hours to just 33 (Whaples, 2001). Other developed countries are even lower. German workers effectively only work 26 hours per week once we take into account annual leave entitlements and public holidays.[2] If these trends continue, we will need to create more jobs to replace these lost hours.

Fifth, we also need to factor in changes in demographics. The number of people seeking employment will surely change. In many developed economies, populations are ageing. If we can fix pension systems, then many more of us may be enjoying retirement, unbothered by the need to work. Indeed, in countries like Japan, there are already significant concerns that there will be too few workers left below retirement age (Hong & Schneider, 2020).

Sixth, we also need to consider how automation will grow the economy. Some of the extra wealth generated by automation will "trickle down" into the economy, creating new job opportunities elsewhere. This argument depends on redistribution mechanisms like taxes which may require adjusting for the new shape of the economy. On the other side, automation may lower costs, making the cost of the basic essentials for living cheaper. If it costs less to live, we may work less.

**9.4 The present**

AI is constantly in the news today. It is impossible to open a newspaper without reading multiple stories about some new application of AI. Many, including those working in the field, are concerned about the impacts it is going to have, especially on jobs. My colleague, Moshe Vardi put it starkly at the 2016 Annual Meeting of the Association for Advancement of Science:

> "We are approaching a time when machines will be able to outperform humans at almost any task… I believe that society needs to confront this question before it is upon us: If machines are capable of doing almost any work humans can do, what will humans do? . . . We need to rise to the occasion and meet this challenge before human labor becomes obsolete." (Rice University, 2016)

There is some evidence that some jobs are starting to be automated, and that some of these jobs are not being replaced by jobs elsewhere. An MIT study from 2017 analysed the impact of automation in the United States from 1993 to 2007 (Acemoglu & Restrepo, 2017). It found that industrial robots reduced overall jobs. Every new robot replaced around 5.6 workers on average. And offsetting gains were not observed in other occupations. In fact, the study estimated that every additional robot per 1,000 workers reduced the total population in employment in the U.S. by 0.34 percent points. Automation also put pressure on the jobs that remained. Every additional robot per 1,000 workers also reduced wages by 0.5 percent. During that 17-year period of the study, the number of industrial robots in the United States quadrupled, eliminating what they estimated was around half a million jobs.

---

[2]https://www.destatis.de/EN/Themes/Labour/Labour-Market/Quality-Employment/Dimension3/3_1_WeeklyHoursWorked.html

The oil industry provides an informative case study of the scale of the challenge. The price of oil collapsed from $115 per barrel in August 2014 to below $30 at the start of 2016[3]. This drove the industry to decrease head-count[4] and introduce more automation. Nearly half a million jobs disappeared from the oil industry worldwide. But now, as the price of oil is rebounding, and the industry is again growing, less than half of those jobs have returned. Automation has reduced the 20 people typically needed to work a well down to just five.[5]

**9.5 Open jobs**

One reason automation won't eliminate some jobs is that, in some cases, automation will just let us do more of that particular job. It's useful in this respect to distinguish between "open" and "closed" jobs. Automation will tend to augment open jobs but replace closed jobs. What do I mean by open and closed jobs?

Closed jobs are those where there is a fixed amount of work. For example, window cleaner is a closed job. There are only a fixed number of windows to be cleaned on the planet, and there's no point cleaning a window that hasn't got dirty again. Window cleaning robots are now starting to appear. Once robots can clean windows cheaper than a human, which I suspect is not far away, the job of human window cleaner will disappear. At least, the job of window cleaner will disappear from developed countries where human window cleaners are expensive and prone to fall off ladders. As a second example, the job of processing insurance claims is a closed job. Customers of an insurance firm only file a certain number of insurance claims. When we automate the processing of insurance claims, we don't generate demand to process more claims. We simply reduce this cost from the insurance industry.

Open jobs by comparison expand as you automate them. For instance, chemistry is an open job. If you are a chemist, AI tools that help automate your job will merely help you do more chemistry. You can push back the frontiers of our understanding of chemistry that much faster. You are unlikely to run out of new chemistry to understand. As a second example, the job of a police detective is an open job. AI tools that help a detective investigate crime will speed up their work, permitting them to consider many of the crimes that are currently ignored.

Of course, most jobs are neither completely open nor completely closed. Take the legal profession. As computers take over more and more routine legal work, the cost of accessing the law will fall. This will expand the market for lawyers, generating more demand and giving all of us better access to legal advice. This will likely create more work for experienced lawyers. But it is hard to imagine that lots of entry level legal jobs will remain. It may be hard for young graduates to compete with robo-lawyers that have read all the legal literature, never need to sleep, never make mistakes and don't need any salary.

---

[3]https://en.wikipedia.org/wiki/2014%E2%80%932016_world_oil_market_chronology;
https://www.statista.com/statistics/262860/uk-brent-crude-oil-price-changes-since-1976/
[4]https://www.statista.com/statistics/465871/global-oil-and-gas-industry-employment-cuts/#statisticContainer
[5]https://www.bloomberg.com/news/articles/2017-01-24/robots-are-taking-over-oil-rigs-as-roughnecks-become-expendable#xj4y7vzkg

**9.6 Partial automation**

One argument put forwards why the automation of 47 per cent of jobs won't translate into mass unemployment is that only some parts of these 47 per cent of job will be automated. This argument is problematic. If you automate parts of a job, then you can usually do the same work with fewer people. Consider again a job like being a lawyer. Closer analysis of the time spent by a lawyer on different aspects of their work suggests that only around one quarter of their time is spent doing tasks that could be automated in the near future (Manyika et al., 2017). Unless we create more legal work (and see the discussion earlier about open and closed jobs for discussion around this topic), averaged over all lawyers, we could therefore do current legal work with three quarters of the current lawyers we have today. Lawyers might lift their game and use the time freed up to do better quality work. But some law firms will simply lower their prices by three quarters, and cut one quarter of their staff to pay for their reduced income.

This argument that only parts of jobs will be automated has even been used to argue that one of the jobs most at risk of automation is, in fact, safe from replacement. Truck drivers need not worry, they say, as there will always be edge cases that defeat the machines. The truck arrives at some engineering works where a road worker signals to the truck by hand. The truck needs to drive around a factory that is not on any GPS maps. Autonomous trucks will simply not be able to cope with such situations. The bad news for truck drivers is that this neglects remote driving. Companies like Starksy Robotics are already testing autonomous trucks in which remote drivers take over when the machine cannot cope. One such remote driver will be able to take care of multiple autonomous trucks. So we might have human driving trucks remotely for some time still, but there will be many fewer of them than now.

**9.7 New jobs**

All technologies create new jobs as well as destroy them. That has been the case in the past, and it might seem reasonable to suppose that it will also be the case in the future. There is, however, no fundamental law of economics that requires the same number of jobs to be created as destroyed. In the past, more jobs were created than destroyed but it doesn't have to be so in the future. This time could be different. In the Industrial Revolution, machines took over many of the physical tasks we used to do. But we humans were still left with all the cognitive tasks. This time, as machines start to take on many of the cognitive tasks too, there's the worrying question: what is left for us humans?

One of my colleagues has suggested there will be plenty of new jobs like robot repair person. I am entirely unconvinced by such claims. The thousands of people who used to paint and weld in most of our car factories got replaced by only a couple of robot repair people. There's also no reason why robots won't be able to repair robots. We already have factories where robots make robots. There are dark factories, factories where there are no people and so no need for lights, in which robots work night and day making other robots. FANUC, one of the largest manufacturers of industrial robots, has operated such a dark factory near Mount Fuji since 2001. This has helped FANUC post annual sales of around $6 billion, selling robots into booming markets like China. Another of my colleagues has suggested we'll have robot

psychologists. As though we'll need one robot psychologist per robot. Robot psychology will be conducted at best by a few people on the planet. So there won't be many jobs looking after the robots.

The new jobs will have to be doing jobs where either humans excel or where we choose not to have machines. Machines might be physically and cognitively better than us at certain jobs, but we nevertheless choose to have humans do them. We might decide that we prefer human judges, hairdressers, influencers or CEOs.

**9.8 The important skills**

Finally, let me address the question of what are the important jobs skills in this race against the machines. The Oxford report identifies three job skills which it is claimed will be difficult to automate in the next few decades: our creativity, our social intelligence, and our ability at perception and manipulation. I would agree with the first two but am uncertain about the third. Computers already perceive the world better than us, on more wavelengths, and at higher precision. What is true is that manipulation is difficult for robots, especially away from the factory floor in uncontrolled environments. This is likely to remain so for some time to come. And even when robots have better manipulation skills than humans, a human may do these manipulation tasks cheaper than an expensive robot.

My best advice here is to head towards one or more of the corners of what I call "the triangle of opportunity". On one corner, we have the geeks, the technically literate. There is a future in inventing the future. It is still very challenging to get computers to program themselves. Computers are also challenged at created a novel future. Be someone then that does that.

Of course, not all of us are technically minded. If you are not, I recommend you head towards one of the other two corners. On one of these corners are those with emotional and social intelligence. Computers are still very poor at understanding emotions. And they don't have emotions of their own. As we will spend more and more of our lives interacting with machines, they will eventually have to understand our emotions better. We may even give them "emotions" of their own so that we can relate to them better. But for some time, computers are likely to have a low emotional intelligence.

On the third and final corner of the triangle of opportunity, we have the creatives and artisans. One reaction to increasing automation in our lives is likely to be an increasing appreciation for that made by the human hand. Indeed, hipster fashion already seems to be embracing this trend. I find it rather ironic then a job like a carpenter, one of the older jobs on the planet, might become one of the safest. So another opportunity is to develop your creativity or learn some artisan skills. Make traditional cheeses. Write novels. Play in a band.

Could computers take on some of these creative tasks? This is a question that has haunted the field of artificial intelligence from the very start. Ada Lovelace famously wrote,

> "The Analytical Engine has no pretensions to originate anything. It can do whatever we know how to order it to perform. It can follow analysis; but it has no power of anticipating any analytical relations or truths. Its province is to assist us to making available what we are already acquainted with."
(Lovelace, 1843, p.722)

Alan Turing attempted to refute Ada Lovelace's objection.

"Who can be certain that 'original work' that he has done was not simply the growth of the seed planted in him by teaching, or the effect of following well-known general principles. A better variant of the objection says that a machine can never 'take us by surprise'. This statement is a more direct challenge and can be met directly. Machines take me by surprise with great frequency."
(Turing, 1950, p. 450)

The multidisciplinary field of "computational creativity" has emerged since Turing's original paper explored this issue. Like AI explores whether computers can model, simulate or replicate human intelligence, computational creativity explores whether computers can model, simulate or replicate human creativity. There is still no consensus on whether machines are creative in the same ways as humans. However, we have seen computers make paintings that have sold for hundreds of thousands of dollars at auction, discover new drugs, and invent devices that have been patented. But even if computers can replace some creative jobs, society may simply choose to value more those things that carry the label "made by hand". Economists would have us believe that the market will respond in this way.

**9.9 Conclusions**

The irony is that our technological future will not be about technology but all about our humanity. The jobs of the future are the very human jobs. AI could help us build a much gentler society. Jobs looking after the young, the sick, the elderly and the handicapped are not, and perhaps never will be, jobs for robots. We need therefore to start valuing them more. By valuing those that look after the young, the sick, the elderly, and the handicapped as much or perhaps even more than those in traditional employment, it will be a more caring society. We might also see a flowering of creativity. This could be the Second Renaissance. Even if robots can create arts or crafts, we will value more objects made by humans. Artificial intelligence could provide the productivity gains that pay for more of us to be artists and artisans.

It is worth remembering past technological changes and learning from history. If we look at the industrial revolution, we made some significant changes to society to deal with the disruption that technological change brought to our lives. We introduced institutions like unions, labour laws, universal education, and the welfare state so that all of us shared the prosperity brought by technological change. We should remember this as we enter another period of profound technological change. Artificial intelligence will change our world dramatically, especially work. We need therefore to think big about the changes to make to society today to ensure it is the world that we want it to be. I shall, however, leave the last word appropriately to AI.

GPT-4: AI will likely disrupt some professions more than others, but the potential for positive change is immense. In some cases, such as with sales, AI will likely lead to a reduction in the overall number of jobs, but will also enable the remaining professionals to become more productive and effective. In other cases, such as with law, AI may lead to an overall improvement in the happiness and satisfaction of the profession.

While it is important to be aware of the ways in which AI may disrupt traditional career paths, it is also important to remember that AI can create new opportunities for growth and advancement. As we move into a future where AI tools are a core part of normal work processes, it is essential for professionals to take advantage of the opportunities that AI presents in order to achieve greater productivity and more meaningful work.
(Hoffman & GPT4, 2023, p. 131)